\documentclass[universe,article,accept,pdftex,moreauthors]{Definitions/mdpi} 
\firstpage{1} 
\pubvolume{1}
\issuenum{1}
\articlenumber{0}
\pubyear{2024}
\copyrightyear{2024}
\externaleditor{Firstname Lastname}
\datereceived{30 October 2024} 
\daterevised{12 December 2024} % Comment out if no revised date
\dateaccepted{17 December 2024} 
\datepublished{ } 
\hreflink{https://doi.org/} % If needed use \linebreak
\newcommand{\Qphi}{$\mathcal{Q}_\phi$}
\def\fdg{\hbox{$.\!\!^\circ$}}
\def\farcs{\hbox{$.\!\!^{\prime\prime}$}}

\Title{Disk Evolution Study Through Imaging of Nearby Young Stars (DESTINYS): Dynamical Evidence of a Spiral-Arm-Driving and Gap-Opening Protoplanet from SAO~206462 Spiral Motion}

\TitleCitation{Disk Evolution Study Through Imaging of Nearby Young Stars (DESTINYS): Dynamical Evidence of a Spiral-Arm-Driving and Gap-Opening Protoplanet from SAO~206462 Spiral Motion}
\Author{Chen Xie $^{1,*}$\orcidA{}, Chengyan Xie $^{2,3}$\orcidB{}, Bin B. Ren $^{4,5,*,\dagger}$\orcidC{}, Myriam Benisty $^{4,5}$\orcidD{}, Christian Ginski $^6$\orcidE{}, Taotao Fang $^3$\orcidF{}, Simon Casassus $^{7,8}$\orcidG{}, Jaehan Bae $^9$\orcidH{}, Stefano Facchini $^{10}$\orcidI{}, Fran\c{c}ois M\'enard $^{11}$\orcidJ{} and Rob G. van Holstein $^{12}$\orcidK{}} 

\AuthorNames{Chen Xie, Chengyan Xie, Bin B. Ren, Myriam Benisty, Christian Ginski, Taotao Fang, Simon Casassus, Jaehan Bae, Stefano Facchini, Fran\c{c}ois M\'enard and Rob G. van Holstein}

\AuthorCitation{Xie, C.; Xie, C.-Y; Ren, B.B.; Benisty, M.; Ginski, C.; Fang, T.; Casassus, S.; Bae, J.; Facchini, S.; M\'enard, F.; van Holstein, R.G.}

\address{
$^{1}$ \quad William H. Miller III Department of Physics and Astronomy, Johns Hopkins University, 3400 N. Charles Street, Baltimore, MD 21218, USA\\
$^{2}$ \quad Lunar and Planetary Laboratory, The University of Arizona, Tucson, AZ 85721, USA; cyxie@arizona.edu \\
$^{3}$ \quad Department of Astronomy, Xiamen University, 1 Zengcuoan West Road, Xiamen 361005, China; fangt@xmu.edu.cn\\
$^{4}$ \quad Observatoire de la C\^{o}te d'Azur, CNRS, Laboratoire Lagrange, Universit\'{e} C\^{o}te d'Azur, Bd de l'Observatoire, CS 34229, F-06304 Nice, %\hl{cedex 4,} %MDPI: Please consider to remove this part.
 France; benisty@mpia.de\\
$^{5}$ \quad Max-Planck-Institut f\"ur Astronomie (MPIA), K\"onigstuhl 17, D-69117 Heidelberg, Germany\\
$^{6}$ \quad School of Natural Sciences, University of Galway, University Road, H91 TK33 Galway, Ireland; christian.ginski@universityofgalway.ie\\
$^7$ \quad Departamento de Astronom\'ia, Universidad de Chile, Casilla %MDPI: Please check if the city name is correct and unify the format of post code for affs 7, 8 and 12.
 36-D, Santiago, Chile; simon@das.uchile.cl\\
$^8$ \quad Data Observatory Foundation, Eliodoro Ya\'n\~ez 2990, Providencia, Santiago, Chile\\
$^9$ \quad Department of Astronomy, University of Florida, Gainesville, FL 32611, USA; jbae@ufl.edu\\
$^{10}$ \quad Dipartimento di Fisica, Universit\'{a} degli Studi di Milano, Via Celoria 16, I-20133 Milano, Italy; stefano.facchini@unimi.it\\
$^{11}$ \quad CNRS, Institut de Plan\'{e}tologie et d'Astrophysique, Universit\'{e} Grenoble Alpes, F-38000 Grenoble, France; francois.menard@univ-grenoble-alpes.fr\\
$^{12}$ \quad European Southern Observatory, Alonso de C\'ordova 3107, Vitacura Casilla 19001, Santiago, Chile; rob.vanholstein@eso.org\\
}

\corres{Correspondence: cx@jhu.edu (C.X.);  bin.ren@oca.eu (B.B.R.)}

\firstnote{Marie Sk\l odowska-Curie Fellow.} 

\abstract{In the early stages of planetary system formation, young exoplanets gravitationally interact with their surrounding environments and leave observable signatures on protoplanetary disks. Among these structures, a pair of nearly symmetric spiral arms can be driven by a giant protoplanet. For the double-spiraled SAO~206462 protoplanetary disk, we obtained three epochs of observations spanning $7$~yr using the Very Large Telescope's SPHERE instrument in near-infrared $J$-band polarized light. By jointly measuring the motion of the two spirals at three epochs, we obtained a rotation rate of $-0\fdg85\pm0\fdg05~{\rm yr}^{-1}$. {This rate corresponds to} a protoplanet at $66\pm3$~au on a circular orbit dynamically driving both spirals. The derived location agrees with the gap in ALMA dust-continuum observations, indicating that the spiral driver may also carve the observed gap. What is more, a dust filament at $\sim$63~au observed by ALMA coincides with the predicted orbit of the spiral-arm-driving protoplanet. This double-spiraled system is an ideal target for protoplanet imaging.}


\begin{document}
\section{Introduction}
Among the over 5000 exoplanets discovered to date, only less than 30 are directly imaged (e.g., \cite{currie23}). While indirect exoplanet detection methods (i.e., radial velocity, transiting) contribute to the majority of the detected exoplanets, these methods are subject to stellar activities (e.g., \cite{lovis10, zhu21}) and thus could only detect exoplanets for ${\gtrsim}1$~Gyr-old stars. In~comparison, directly imaged planets are around young systems that could be ${\lesssim}10$~Myr old (e.g., \cite{keppler18, haffert19}). The~difference in the mass-period distribution of detected exoplanets in systems with different evolutionary stages (e.g., \cite{bae18}) calls for more directly imaged exoplanets to complement our understanding of planetary system~evolution.

While existing high-contrast imaging surveys of exoplanets over the past decade have low detection rates for exoplanets (e.g., \cite{nielsen20, vigan21}), targeted campaigns are becoming more successful in imaging stellar and planetary companions at high contrast (e.g., \cite{bohn20, franson23, currie23b, matthews24}). However, targeted campaigns infer planetary existence based on stellar signal deviations (e.g., astrometry, radial velocity), and~this could require decades of monitoring of stars. Around young stars that were not monitored due to increased stellar activities, young planets can gravitationally interact with their birthplaces---the protoplanetary disks, and~leave observable subsignatures (e.g., spirals, gaps) on these surrounding environments (e.g., \cite{dong17, dong15planet, bae16}). With~some assumptions, we can infer planetary existence---and even their mass, location, and~orbit -- based on substructures created by planet-disk interaction (e.g., \cite{dong15planet, bae16, teague18}).

Among the planet-induced subsignatures, a~pair of nearly symmetric spiral arms can be driven by a single $5$--$10M_{\rm Jup}$ companion on a wide orbit \citep{dong15planet, bae16, dong18}. With~the capabilities of the current generation of high-contrast imagers, the~age, mass, and~orbits of these planets make them the best targets for planet imaging \citep{bae18}. In~fact, based on the co-motion between an M-star and a spiral,  Ref.~\cite{xie23} validated the theory of companion-driven spirals. This provides both theoretical and observational support to locate spiral-arm-driving planets based on spiral motion measurements~\cite{ren20}. 

The current generation of high-contrast imagers has resolved nearly two dozen spiral arm systems in the past decade (e.g., \cite{benisty15, wagner15, shuai22}). With~polarized light observations offering the best quality images for morphological analysis (e.g., \cite{monnier19}), we can obtain precise motion for the spiral arms with images spanning ${\sim}5$-yr or mor (e.g., \cite{ren20, xie21}). The~measured motion rate (i.e., angular speed) allows us to distinguish the leading motion and thus formation mechanisms for spiral arms~\cite{dong18, ren20}. Specifically, if~a spiral is driven by a companion, it is expected to rotate entirely as a rigid body at an identical rate as the driver. This thus allows the prediction of the orbit of these hidden planetary drivers (e.g., \cite{ren20, xie21, ren23v1247, safonov22}) for follow-up targeted imaging efforts (e.g., \cite{wagner19, cugno24}). However, despite the validated theory of companion-disk interaction (i.e., HD~100453: \cite{xie23}), none of the planetary driver candidates have been~confirmed.

Spiral motion rates can guide targeted high-contrast imaging to further find and characterize the spiral-arm-driving planets. This requires accurate and precise spiral motion rates to guide observation design for high-contrast imaging. To~best characterize the motion rates, we need to observe the spirals at different temporal epochs, ideally with identical data quality (e.g., wavelength, instrument setup, exposure time) \cite{ren20}. Among~existing spiral arm systems, e.g.,~\cite{shuai22}, only one system has derived motion rates consistent with being driven by a single planet \cite[i.e., MWC~758:][]{ren20}. While there is an ongoing survey to re-observe known spiral systems and characterize their motion \citep{ren23v1247}, it is necessary to explore the stability of the preliminary spiral motion rates using observations with longer temporal~separations.

With an estimated age of $12_{-6}^{+4}$~Myr and stellar mass of $1.6\pm0.1M_\odot$ \cite{garufi18},  the~SAO~206462 (i.e., HD~135344~B) system is located at $135.0\pm0.4$~pc~\cite{GaiaDR3}. The~scattered light asymmetry in 2009 \textit{Hubble Space Telescope} near-infrared imaging observations in Ref.~\cite{grady09} was resolved to be two spiral arms from ground-based high-contrast imaging~\cite{muto12, garufi13}. Based on the morphology of the spirals, they can be explained by different mechanisms. On~the one hand, they can be excited by two individual planets: $0\farcs39$ and $0\farcs9$ (or $53$~au and $120$~au) \cite{muto12}, or~two planets at $21$~au and $23$~au (if they are inside the scattered light cavity) or at $95$~au and $162$~au (if they are exterior to) \cite{stolker16b}. On~the other hand, they can be driven by a single planet at $100$--$120$~au~\cite{bae16, dong17}. Alternatively, gravitationally unstable protoplanetary disks can also generate a pair of spirals~\cite{dong15gi}. What is more, spirals in protoplanetary disks can also be excited by recent stellar flybys~\cite{cuello19}, yet for SAO~206462 this has been ruled out using the stellar location and motion rates from \textit{Gaia}~DR3~\cite{shuai22}. 

While the attempts in explaining the spiral morphology in scattered light provided inconclusive results, tracking the motion of the spirals can dynamically test leading spiral formation mechanisms~\cite{ren20, xie23}. The~motion rates of the SAO~206462 spirals were constrained using observations with a $1$-yr separation, which ruled out gravitational instability mechanism for the spirals~\cite{xie21}. While the motion rates helped investigate the nature of a candidate~\cite{cugno24}, the~$1$-yr separation still resulted in two possible planetary explanations: the two spirals can be either driven by one single planet at ${\sim}90$~au, or~two individual planets at ${\sim}120$~au and ${\sim}50$~au. In~the former scenario, a~$5$--$15M_{\rm Jup}$ planet, e.g.,~\cite{bae16} would be ideal for targeted high-contrast imaging using the current generation of instruments. In~the latter scenario, the~planets can be ${\lesssim}0.5M_{\rm Jup}$ \cite[][]{muto12, stolker16b} which is best accessible only with next-generation instruments expected later in this decade, e.g.,~\cite{currie23}. Here with the Disk Evolution Study Through Imaging of Nearby Young Stars (DESTINYS, PI: C.~Ginski;~\cite{destinys1}) survey, we re-observed SAO~206462 in 2022 to establish a $7$~yr temporal separation. With~the observations, we here characterize the motion of the two spirals, and~guide follow-up targeted high-contrast imaging on the spiral-arm-driving planet(s).

%%%%%%%%%%%%%%%%%%%%%%%%%%%%%%%%%%%%%%%%%%
\section{Observation and Data~Reduction}
We collected SAO~206462 observations at three epochs (i.e., in~2015, 2016, and~2022) that span $7$~yr for spiral motion analysis. All observations were carried out using the infrared dual-band imager and spectrograph (IRDIS;~\cite{dohlen08}) on the SPHERE instrument~\cite{beuzit19} at the Very Large Telescope (VLT). The~observations were under the dual-beam polarimetric imaging (DPI;~\cite{irdap2}) mode to help reveal the SAO~206262 spiral arms in polarized light. To~establish the longest temporal separation of high-quality data for analysis, and~reduce the possibility that wavelength-dependent imaging manifests spurious spiral motion, all the SAO~206462 observations for analysis here are in $J$-band (1.245~{\textmu}m).

\textls[-15]{The archival observations in 2015 and 2016 are already presented in~\cite{stolker16b, stolker17}. Among these} observations, we chose the 3 May 2015 (under ESO program 095.C-0273) and 4 May 2016 (under ESO program 097.C-0885) observations identified by~\cite{xie21} to have the best image quality for spiral morphology determination and motion analysis. In~the 1 h 42 m observation block in 2015, there were 50 exposures, and~each exposure had 3 integrations of \mbox{32 s,} totaling 4800s on-target time. In~the 1 h 58 m observation block in 2016, there were 92 exposures, and~each exposure had 2 integrations of 32 s, totaling 5888s of on-target~time.

%http://localhost:8888/notebooks/Documents/saffron/SAO206462/plot_paper/plot3.ipynb
%http://localhost:8888/notebooks/Documents/saffron/SAO206462/plot_paper/plot3-inc16.74.ipynb

We observed SAO~206462 using VLT/SPHERE/IRDIS in $J$-band DPI mode under ESO program 1104.C-0415 in the DESTINYS survey (\cite{destinys1}, PI: C.~Ginski). The~observation on UT 30 March 2022  was from UT 04:18:45 to UT 05:09:47. Among~this 0 h 51 m observation block, we had 64 exposures, and~each exposure had 1 integration of 32 s, totaling 2048s on-target~time.

\begin{figure}[H]
\includegraphics[width=\textwidth]{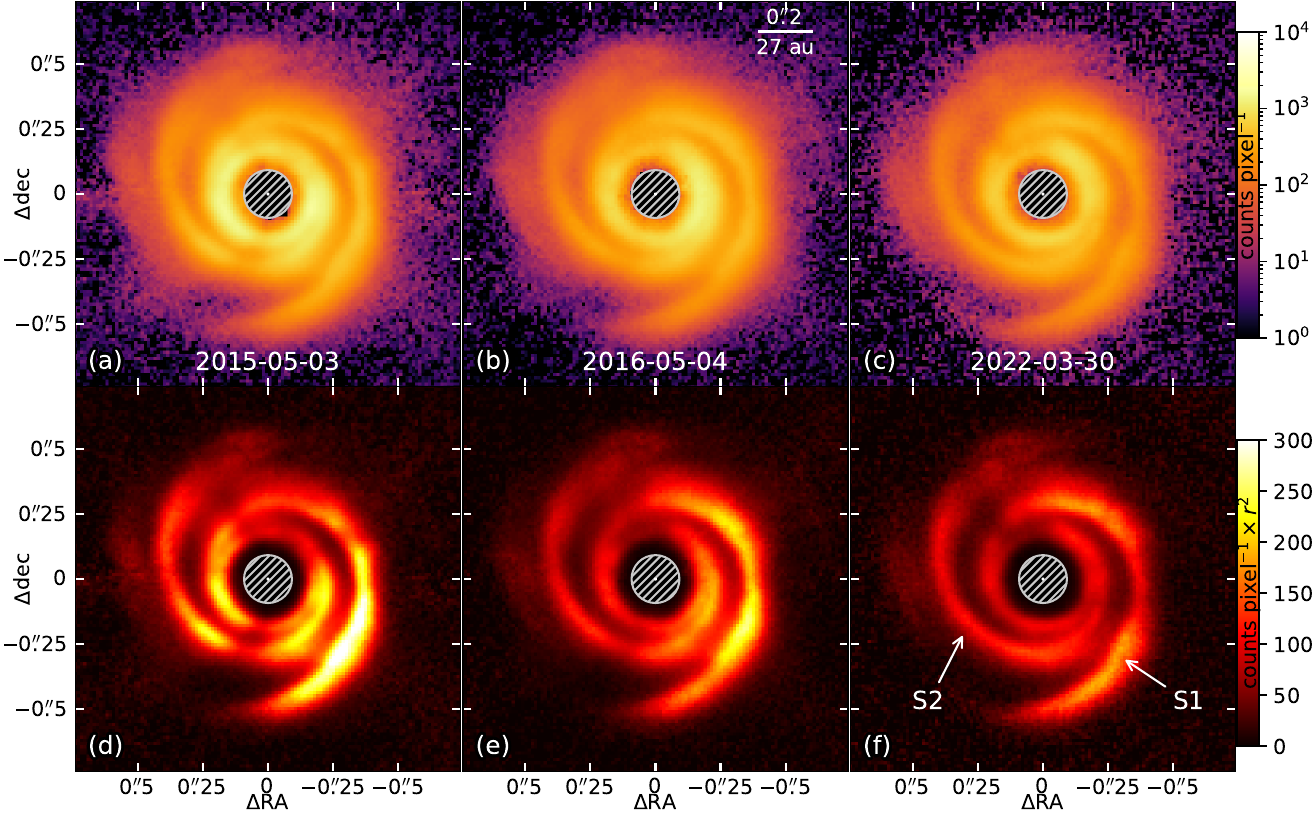}
\caption{Three-epoch images of SAO~206462 spiral arms in $J$-band polarized light. \textit{Top}: \Qphi\ maps with stellar polarization removed, with~the surface brightness (SB) distribution displayed in log scale. \textit{Bottom}: Images from \Qphi\ maps with $r^2$-scaling, which is normalized at $0\farcs5$ and displayed in linear scale. Note: the $r^2$-scaled maps are shadowed by the SAO~206462 inner disk(s) \cite{stolker16b, stolker17}, and~thus they do not map the actual surface density distribution of dust particles scattering light in $J$-band. (Note: the data are available at the CDS.)} \label{fig1}
\end{figure}

We reduced the three IRDIS DPI datasets using the \texttt{IRDAP} pipeline from Ref.~\cite{irdap2} that was modified by Ref.~\cite{ren23}. To~prepare the datasets for analysis, \texttt{IRDAP} performs alignment, bad pixel correction, and~sky background subtraction. It then processes the datasets using polarimetric differential imaging (PDI). By~adopting the default \texttt{IRDAP} reduction parameters, we use median-combined output \Qphi\ images with stellar polarization removed for spiral motion analysis, see Figure~\ref{fig1}. These \Qphi\ images trace the surfaces of the SAO~206462 protoplanetary disk, e.g.,~\cite{monnier19} for spiral morphology and motion analysis, e.g.,~\cite{ren20}.

On protoplanetary disk surfaces, the~intensity of incident starlight decreases as $r^{-2}$ with $r$ being the stellocentric distance. As~a result, the~light scattered by dust particles in the disk -- when observed by our telescope -- reduces intensity accordingly. For~spiral morphology and motion analysis, we rescaled the brightnesses of the \Qphi\ images by multiplying them with a $r^2$ map -- assuming a {$16\fdg7$} inclination from face-on and a $61\fdg9$ position angle for the major axis of the disk~\cite{bohn22} -- to obtain the $r^2$-scaled maps (normalized at $0\farcs5$) in Figure~\ref{fig1}. 

%%%%%%%%%%%%%%%%%%%%%%%%%%%%%%%%%%%%%%%%%%
\section{Results and~Discussion}
\unskip

\subsection{Results}\label{sec-result}
We followed Ref.~\cite{ren20} to combine the three epochs of SAO~206462 observations for spiral arm location and motion analysis. Given the low disk inclination, we assumed an infinitely thin disk for subsequent analysis in Section~\ref{sec-result}.

%http://localhost:8888/notebooks/Documents/2024Fall/SAO206462/inc16deg/2.a%20fit%20motion%203%20epochs%20SINGLE.ipynb

\begin{figure}[H]
\includegraphics[width=0.95\textwidth]{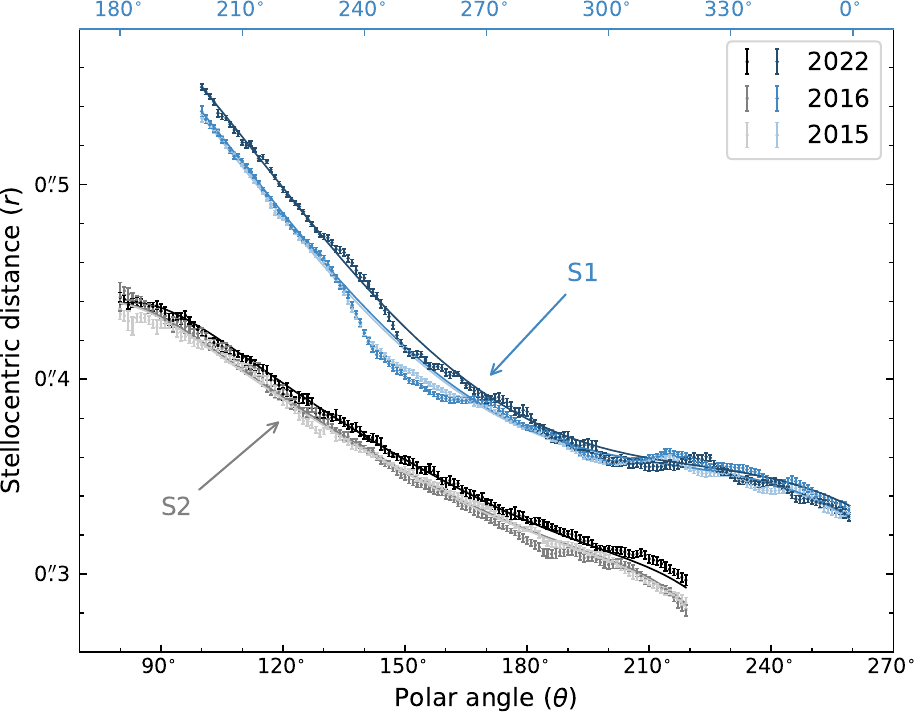}
\caption{Three-epoch spiral arm locations and single driver motion SAO~206462. The~lines are from the best-fit $0\fdg85$~yr$^{-1}$ counter-clockwise motion for both spirals, the~associated $1\sigma$ error is $0\fdg05$~yr$^{-1}$. For~a $1.6\pm0.1M_\odot$ star, the~spiral rotation rate corresponds to a circular driver orbit of $66\pm3$~au, or~$0\farcs49\pm0\farcs02$ from the~star. } \label{fig-motion}
\end{figure}

First, for~each arm, we obtained the radial locations of the spiral peaks at different epochs, e.g.,~\cite{ren23v1247}. We transformed the $r^2$-scaled maps in Figure~\ref{fig1} to polar coordinates. For~each angular location, which is defined as a clockwise separation from the major axis of the disk, we fit a normal distribution to obtain the peak and its associated error on the spiral radial profile. With~a $1^\circ$ angular step, we present the peaks and their errors \endnote{The uncertainties in this paper are $1\sigma$ unless otherwise specified.} in Figure~\ref{fig-motion}.

Second, for~the peaks, we use $p$-degree polynomials ($p\in\mathbb{N}$), \[r = \sum_{i=0}^{p} c_i \theta^i,\] to simultaneously derive the spiral morphological parameters and its pattern motion, see Figure~\ref{fig-motion}. For~each arm at the three epochs, we followed Ref.~\cite{ren20} to fit an identical but angularly offset polynomial to obtain the best-fit in Figure~\ref{fig-motion} using dummy variables; the best-fit offset will be the angular rotation among the epochs. Following Ref.~\citep{ren23v1247}, the~polynomial fitting uncertainty will be propagated with three epochs of $0\fdg08$ SPHERE true north pointing uncertainty in Ref.~\cite{maire16} and three epochs of SPHERE astrometric uncertainty of $0\fdg16$. The~astrometric accuracy of SPHERE/IRDIS is $\sim$2~mas ($\sim$0.16~pixels; \citep{Zurlo2014}), thus translating into $0\fdg16$ in polar~coordinates. 

Third, we obtained the spiral motion rates under two planet-driven scenarios in Ref.~\cite{ren20}, and~positive rotation rates correspond to clockwise rotation here. On~the one hand, we fit the two spirals independently. For~S1, the~best-fit motion is $-0\fdg84\pm0\fdg05$~yr$^{-1}$ with $\chi_{v}^{2}$ of 3.49 using a $6$-degree polynomial; for S2, it is $-0\fdg87\pm0\fdg05$~yr$^{-1}$  with $\chi_{v}^{2}$ of 1.83 using a $8$-degree polynomial. The~spiral motions of S1 and S2 are consistent within $1\sigma$, suggesting that they are driven by a single driver. On~the other hand, we fit the two spirals simultaneously, and~the best-fit motion rate is
\begin{equation}\label{rate}
\omega = -0\fdg85\pm0\fdg05~{\rm yr}^{-1} 
\end{equation} for two $7$-degree polynomials with $\chi_{v}^{2}$ of 2.67, see the lines in Figure~\ref{fig-motion}. For~the motion of the two spirals, we do not further test gravitational instability (e.g., \citep{Lodato2005}) as the spiral formation mechanism, since it has been confidently ruled out with the study on $1$-yr separation for SAO~206462~\cite{xie21}. Although~the disk images have intensity variations possibly due to inner disk shadowing~\cite{bohn22}, motion measurements using different parts of the disk (i.e., S1, S2, and~two arms combined) are consistent with each other, indicating the limited impact of the intensity variation in the motion measurement of SAO~206462.

\subsection{Discussion}
\subsubsection{Implications}
\begin{figure}[H]
\includegraphics[width=0.45\textwidth]{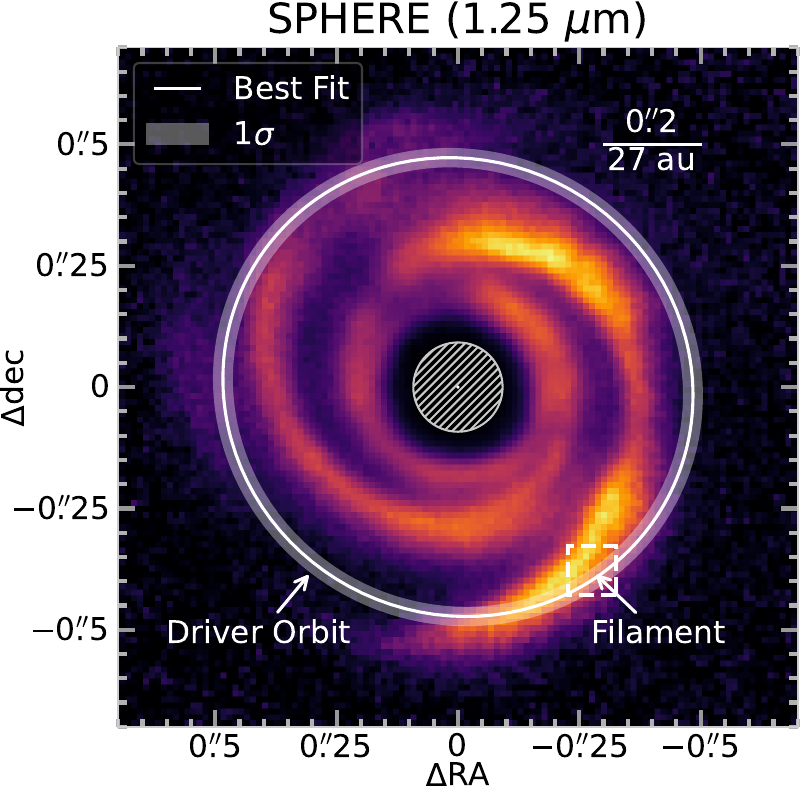}
\includegraphics[width=0.45\textwidth]{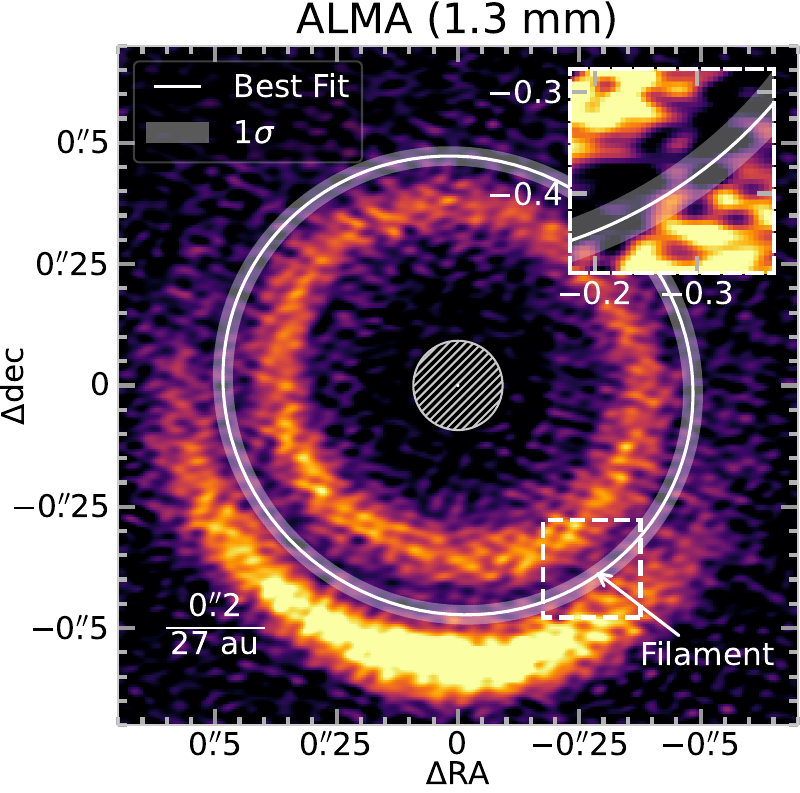}
\caption{Comparison between the SAO~206462 spiral driver orbit and the disk detected in scattered light and thermal emission. \textit{Left:} SAO~206462 spiral driver orbit of $66\pm3$~au, or~$0\farcs49\pm0\farcs02$, based on three-epoch double-spiraled motion. The~orbits are overplotted on the $r^2$-scaled map from Figure~\ref{fig1}f. The~Ref.~\cite{casassus21} dust filament at $(-0\farcs276, -0\farcs378)$ in the dashed white box could trace the spiral-arm-driving protoplanet. \textit{Right:} ALMA continuum image of SAO~206462 from Ref.~\cite{casassus21}, showing the disk gap and a filament (dashed white box) connecting the inner and outer rings. The~inset zooms on the dust filament that intersects with the predicted orbit from spiral~motion.} \label{fig-orbit}
\end{figure}

For the double-spiral driver in SAO~206462, its orbital period is \begin{equation}
    T = 424 \pm 25~{\rm yr}
\end{equation} from Equation~\eqref{rate} assuming a circular orbit. The~corresponding stellocentric radius is $66\pm3$~au (i.e, $0\farcs49\pm0\farcs02$) for a $1.6\pm0.1M_\odot$ star, see the circular orbit in Figure~\ref{fig-orbit}. In~comparison with direct imaging candidates, {our data suggest} that the \textit{James Webb Space Telescope} (\textit{JWST})/NIRCam candidate in Ref.~\cite{cugno24} is not driving the two spirals at ${>}5\sigma$.

The predicted driver orbit coincided with the disk gap observed in the thermal emission and a filament structure that connects two rings, see Figure~\ref{fig-orbit}. The~orbit of the spiral driver agrees with the disk gap, which indicates the possibility of a companion driving spirals while carving a millimeter gap, similar to the case of V1247~Ori \citep{ren23v1247}. The~filament observed in Ref.~\cite{casassus21} is located at ($\Delta$RA, $\Delta$dec) = $(-0\farcs276, -0\farcs378)$, or~$0\farcs47$ from the star (Figure~\ref{fig-orbit}). This suggests that the filament coincides with the orbit of the spiral-driving protoplanet in Figure~\ref{fig-orbit} that drives the two spirals simultaneously. In~combination with the orbit derived from spiral motion, the~disk gap, and~the dust filament, we can {predict} the possible location of the hidden protoplanet that creates those~features. 

While this protoplanet is currently not detected, the~existence of dust particles causing significant extinction in near-infrared wavelengths makes it more challenging to image this spiral-arm-driving protoplanet. To~image this spiral-arm-driving companion, high spectral resolution observations using single-mode fibers (e.g., KPIC; \cite[][]{delorme21} and HiRISE; \cite[][]{Vigan2024}) might help confirm the existence of this protoplanet by directly probing its atmospheric compositions, e.g.,~\cite{delorme21}. However, these observations would need high-precision locations for the protoplanet within the effective response of a single mode fiber, making it challenging to efficiently schedule the observations unless with multiple pointing~attempts.

\subsubsection{Disk~Flaring}
Disk flaring will affect the deprojection of the observed disk $r^2$-scaled map and thus affect the measurements of spiral arm location and motion analysis, especially for disks with high inclinations (i.e., ${>}30^\circ$; \cite{xie23}). In~Section~\ref{sec-result}, we assumed a thin disk because of the low disk inclination of SAO~206462. To~examine the impact of disk flaring on motion measurement for SAO~206462, we adopted a disk model of $h=0.048\times d^{1.15}$ from Ref.~\citep{Andrews2011}, where $h$ represents the disk scale height and $d$ is the distance in cylindrical coordinates. Then we deprojected the disk surface map and performed the same measurements as in Section~\ref{sec-result}. If~we fit the two spirals simultaneously, the~best-fit motion is $-0\fdg82\pm0\fdg05$~yr$^{-1}$ with $\chi_{v}^{2}$ of 2.89 using a $6$-degree polynomial. If~we fit the two spirals independently, the~best-fit motions for S1 and S2 arms are $-0\fdg80\pm0\fdg05$~yr$^{-1}$ and $-0\fdg92\pm0\fdg05$~yr$^{-1}$ with $\chi_{v}^{2}$ of 1.89 and 3.85, respectively. 

After considering the disk flaring, the~spiral motions of S1 and S2 are still consistent within $1.3\sigma$. In~fact, if~the two spirals are independently driven by two planets, the~orbits for the spiral arm motion rate would be $69\pm3$~au for S1, and~$62\pm3$~au for S2. While the two planets are mathematically close to each other, they are not under low order mean motion resonance, which would make the system unstable, e.g.,~\cite{wang18}, consistent with the single driver scenario where two spirals have the same pattern motion as in Section~\ref{sec-result}.

\subsubsection{Limitations}
While the study using $1$-yr separation in Ref.~\cite{xie21} was able to rule out the gravitational instability induced motion, the~companion-driven rates still had large uncertainties, raising the question of whether the two spirals are driven by a single or two individual drivers. In~fact, spiral motion studies have been assuming independent azimuthal angles in Figure~\ref{fig-motion}, e.g.,~\cite{ren20, xie21}, and~thus it is necessary to take into account the correlation among adjacent pixels. Before~such studies are available, we can still use this short-then-long epoch approach to investigate the uncertainties in a physically motivated~way.

\section{Conclusions}
With three epochs of observations for the SAO~206462 (i.e., HD~135344~B) protoplanetary disk hosting two spiral arms in near-infrared scattered light, we have measured the $7$~yr motion of the two spirals. For~a $1.6\pm0.1M_\odot$ star and assuming a circular orbit, the~obtained $\omega = -0\fdg85\pm0\fdg05~{\rm yr}^{-1}$ rotation rate corresponds to a hidden planet at $66\pm3$~au, consistent with the disk gap in ALMA dust-continuum observations. For~a star located at $135.0\pm0.4$~pc, it corresponds to an angular separation of $0\farcs49\pm0\farcs02$ from the~star.

The dust filament in millimeter observations in Ref.~\cite{casassus21} coincides with the orbit of the spiral-arm-driving planet, see Figure~\ref{fig-orbit}. If~that filament is at the exact or approximate location of a forming giant protoplanet, that planet could be responsible for driving the two spiral arms and carving the disk gap in SAO~206462.

With Ref.~\cite{xie23} validating the companion-disk interaction theory for an M star exciting spiral arms in the HD~100453 system, now we have two double-spiraled systems suggesting a single planetary driver for both MWC~758 from~\cite{ren20} and SAO~206462 from this study. Given that spiral-arm-driving giant protoplanets have not been confirmed with direct imaging, and~with the demonstrated sensitivity of \textit{JWST} \cite{carter23}, these planets are ideal targets for current space-based exoplanet imaging since they can minimize the contamination of dust signals from protoplanetary disks using multi-wavelength imaging~observations.

%%%%%%%%%%%%%%%%%%%%%%%%%%%%%%%%%%%%%%%%%%
\vspace{6pt} 

\authorcontributions{Conceptualization, B.B.R. and C.X. (Chen Xie); methodology, B.B.R. and C.X. (Chen Xie); validation, C.X. (Chen Xie), C.X. (Chengyan Xie) and B.B.R.; formal analysis, C.X. (Chen Xie), C.X. (Chengyan Xie) and B.B.R.; software, B.B.R., C.X. (Chen Xie) and C.X. (Chengyan Xie); project administration, B.B.R.; supervision, B.B.R.; resources, M.B., C.G. and B.B.R.; writing---original draft preparation, B.B.R., C.X. (Chen Xie) and C.X. (Chengyan Xie); writing---review and editing, C.X. (Chengyan Xie), M.B., C.G., T.F., S.C., J.B., S.F., F.M., R.G.v.H.; visualization, C.X. (Chen Xie) and B.B.R.; funding acquisition, M.B., B.B.R., T.F. and F.M. All authors have read and agreed to the published version of the manuscript.}

\funding{T.F.~and C.-Y.X. (Chengyan Xie)~are supported by the National Key R\&D Program of China No.~2017YFA0402600, project S202010384487 XMU Training Program of Innovation and Enterpreneurship for Undergraduate, and~NSFC grants No.~11525312, 11890692. This project has received funding from the European Research Council (ERC) under the European Union's Horizon 2020 research and innovation programme (PROTOPLANETS, grant agreement No.~101002188). This project has received funding from the European Research Council (ERC) under the European Union's Horizon Europe research and innovation programme (Dust2Planets, grant agreement No.~101053020). This research has received funding from the European Union's Horizon Europe research and innovation programme under the Marie Sk\l odowska-Curie grant agreement No.~101103114.}

\dataavailability{The original data presented in the study are openly available at the CDS via anonymous ftp to \url{cdsarc.u-strasbg.fr} (\url{130.79.128.5}) or via \url{https://cdsarc.cds.unistra.fr/viz-bin/cat/J/other/Univ/}.}

\acknowledgments{We thank Valentin Christiaens for discussion. Based on observations collected at the European Organisation for Astronomical Research in the Southern Hemisphere under ESO programmes 095.C-0273, 097.C-0885, and 1104.C-0415. This paper makes use of the following ALMA data: ADS/JAO.ALMA\#2018.1.01066.S. ALMA is a partnership of ESO (representing its member states), NSF (USA) and NINS (Japan), together with NRC (Canada), NSTC and ASIAA (Taiwan), and KASI (Republic of Korea), in cooperation with the Republic of Chile. The Joint ALMA Observatory is operated by ESO, AUI/NRAO and NAOJ. This work has made use of data from the European Space Agency (ESA) mission {\it Gaia} (\url{https://www.cosmos.esa.int/gaia}), processed by the {\it Gaia} Data Processing and Analysis Consortium (DPAC, \url{https://www.cosmos.esa.int/web/gaia/dpac/consortium}). For this Special Issue ``New Insights into High-Energy Astrophysics, Galaxies, and Cosmology--Celebrating the 10th Anniversary of the Re-establishment of the Department of Astronomy at Xiamen University (2012–2022)'' of \textit{Universe}, B.B.R., C.X., and C.-Y.X.~are Class of 2013, 2016, and 2022 alumni from Xiamen University, respectively. We are grateful for the education, trainings, and resources received during our undergraduate studies.}

\conflictsofinterest{The authors declare no conflicts of~interest.} 

%%%%%%%%%%%%%%%%%%%%%%%%%%%%%%%%%%%%%%%%%%
\begin{adjustwidth}{-\extralength}{0cm}
\printendnotes[custom]
\reftitle{References}

\PublishersNote{}
\end{adjustwidth}
\end{document}